# Some aspects of the Hadamard's ill-posedness in the hydrodynamical problem

by

*Michele Romeo*[*]

---

## Abstract

Navier-Stokes equations establish the hydrodynamical problem by definition. The importance of these equations is quite natural to understand if we focus on the role they assume in a large spectrum of dynamical problems which involve 'fluids'. Neverthless, they are an undeniable source of pure mathematical problems in PDE's theory [Ref0]. The essential core of their formulation was primarily well structured on the simple concept that the infinitesimal portions of a continuous medium, which flows locally in some manner, must obey in a 'bounded' domain to the same fundamental rules we use to describe the evolution of isolated lagrangian systems, basically momentum and mass conservation laws, so that the consequent architecture of the mathematical implant appears very clear and understandable. Looking to the framework of the numerical solvers, taking in care the richness of their differential structure and the correlated existence of complex dynamics which are mathematically coherent, I try to put in light in the most simple way the fundamental difficulty that arises when we have to impose a reasonable 'initial values problem' in order to simulate numerically well known fluid-dynamical scenarios, trying at the same time to offer a possible method to avoid such an obstacle in determining simulation parameters from which starting in respect of the essential Hadamard's point of view of the Cauchy problem.

---

[*] Lehrstuhl für Aerodynamik, Technische Universität München
   michele.romeo.mr@gmail.com

**Introduction - mathematical background**

In order to determine the physical behaviour of several fundamental problems in Hydrodynamics [Ref1], it is well known that the essential Navier-Stokes mathematical model [Ref9] has a sufficiently strong differential structure to impose the right dynamical constraints on the 'fluid scenario' we have to consider. As it follows, we report the basic vector model in a cartesian inertial frame of reference, by looking it for general compressible newtonian fluids (the real ones):

$$\rho\left(\frac{\partial \vec{u}}{\partial t} + \vec{u} \cdot \nabla \vec{u}\right) = -\nabla p + \nabla \cdot \left(\mu\left(\nabla \vec{u} + (\nabla \vec{u})^T\right)\right) + \nabla(\lambda \nabla \cdot \vec{u}) + \rho \vec{f} \quad [1]$$

$$\frac{\partial \rho}{\partial t} + \nabla \cdot \rho \vec{u} = 0 \quad [2]$$

where

$$\vec{u} = (u(x,y,z,t), v(x,y,z,t), w(x,y,z,t))$$

is the fluid velocity field, $p$ is the dynamic pressure, $\rho$ is the fluid density (generally depending on space and time), $f$ is a generic external force (like gravity), $\mu$ is the dynamic viscosity and $\lambda$ is related to a term which produces a viscous effect associated with the volume change (named 'bulk viscosity'). Sometime $\mu$ and $\lambda$ are given fixed in space and time but the very general framework in which usuallly we move by going through mesoscale hydrodynamics (like taking in care interfacial or high-gradient density effects in multiphase fluids) needs to consider both of them depending on space through one-point functions.
The formula [1] states the conservation of momentum while the expression [2] states basically the conservation of mass at infinitesimal level. The complete set of differential equations [1+2] is named 'Navier-Stokes hydrodynamical system' and, as it shows, it bases its rightness on the fundamental assumption of 'description of a fluid like a deformable continuum'.
It is clearly a strongly nonlinear model whose primitive form stems from the *Reynolds transport theorem* [Ref2]; when we expand the $\nabla$-operator and extract the single-component behaviour, it leads to the following sub-set of nonlinear PDEs:

$$\rho\left(\frac{\partial u}{\partial t} + u\frac{\partial u}{\partial x} + v\frac{\partial u}{\partial y} + w\frac{\partial u}{\partial z}\right) = -\frac{\partial p}{\partial x} + \frac{\partial}{\partial x}\left(2\mu\frac{\partial u}{\partial x} + \lambda \nabla \cdot \vec{u}\right) + \frac{\partial}{\partial y}\left(\mu\left(\frac{\partial u}{\partial y} + \frac{\partial v}{\partial x}\right)\right) + \frac{\partial}{\partial z}\left(\mu\left(\frac{\partial u}{\partial z} + \frac{\partial w}{\partial x}\right)\right) + \rho f_x$$

$$\rho\left(\frac{\partial v}{\partial t} + u\frac{\partial v}{\partial x} + v\frac{\partial v}{\partial y} + w\frac{\partial v}{\partial z}\right) = -\frac{\partial p}{\partial y} + \frac{\partial}{\partial x}\left(\mu\left(\frac{\partial v}{\partial x} + \frac{\partial u}{\partial y}\right)\right) + \frac{\partial}{\partial y}\left(2\mu\frac{\partial v}{\partial y} + \lambda \nabla \cdot \vec{u}\right) + \frac{\partial}{\partial z}\left(\mu\left(\frac{\partial v}{\partial z} + \frac{\partial w}{\partial y}\right)\right) + \rho f_y \quad [3]$$

$$\rho\left(\frac{\partial w}{\partial t} + u\frac{\partial w}{\partial x} + v\frac{\partial w}{\partial y} + w\frac{\partial w}{\partial z}\right) = -\frac{\partial p}{\partial z} + \frac{\partial}{\partial x}\left(\mu\left(\frac{\partial w}{\partial x} + \frac{\partial u}{\partial z}\right)\right) + \frac{\partial}{\partial y}\left(\mu\left(\frac{\partial w}{\partial y} + \frac{\partial v}{\partial z}\right)\right) + \frac{\partial}{\partial z}\left(2\mu\frac{\partial w}{\partial z} + \lambda \nabla \cdot \vec{u}\right) + \rho f_x$$

where we have usefully retained the divergence of the velocity.
Looking to the left hand side of the equation [1], it is quite clear that nonlinearity is strongly related to the compressibility effects (e.g. see following figure) and it is well known that it is crucial to simulate turbulent phenomena (one of the main goals of the hydrodynamic modeling), that is the main reason why we consider the compressible basic model of Navier-Stokes given for the Newtonian case,

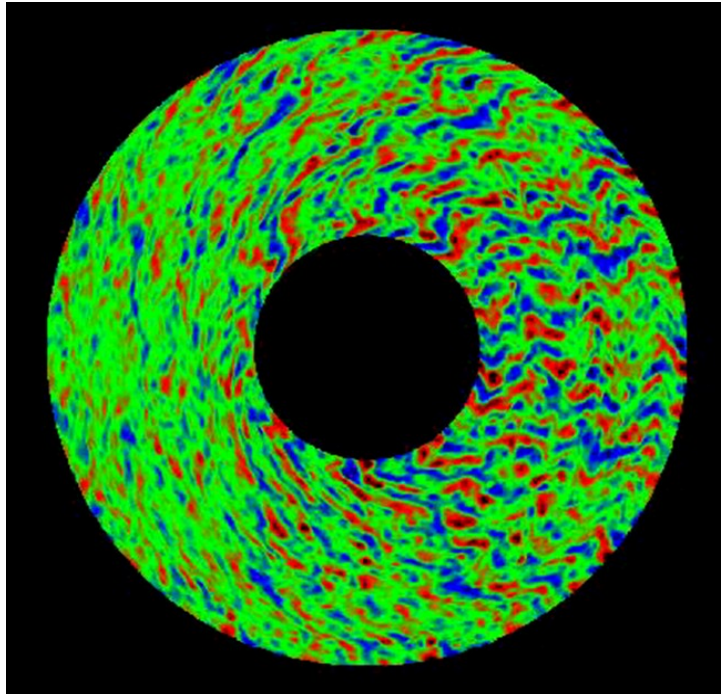

Closeup of poloidal cross-section in a tokamak.
It illustrates the nonlinear generation of zonal flows.

However, many simple real cases are in respect of the incompressible model.
In any case, when we have to reproduce the physical behaviour of a generic fluid problem, almost always we can base finite-difference modeling for weakly compressible fluids on the fundamental differential structure of the Navier-Stokes equations, keeping in care the specific 4D-shape of the fluid scenario by establishing the mathematical-related boundary problem togheter with the initial-value problem of interest.
Nevethless, in most cases we can ignore the boundary problem without loss of significance because the need to put in light the local behaviour in a fluid sub-domain **Ω** which is far enough from the boundaries of the fluid domain **Σ** (physical boundaries do not affect the local behaviour).
In such a case the hydrodynamical problem we have to solve reduces to a local *Cauchy problem* as it follows,

$$\rho\left(\frac{\partial \vec{u}}{\partial t} + \vec{u} \cdot \nabla \vec{u}\right) = -\nabla p + \nabla \cdot \left(\mu\left(\nabla \vec{u} + \left(\nabla \vec{u}\right)^T\right)\right) + \nabla\left(\lambda \nabla \cdot \vec{u}\right) + \rho \vec{f}$$

$$\frac{\partial \rho}{\partial t} + \nabla \cdot \rho \vec{u} = 0 \qquad [4]$$

$$\vec{u}(x,y,z;0) = \vec{h}(x,y,z) \quad \forall (x,y,z) \in \Omega \subset \Sigma \subset \mathbb{R}^3$$

in which the only boundary of interest is the immersed **∂Ω**, the unphysical one (or the boundary of the sub-domain), and the related boundary problem is automatically solved from the bulk behaviour by the viscous effects (since it is possible to assume that ∂Ω behaves like a *membrane*, an immersed boundary problem is well stated). It is known that the validity of the model [4] is very general because the large diffusion of Newtonian-like fluids in Hydrodynamics.

Anyway, sometime we have to take in count, as stated above, just small given perturbations $\varepsilon_\rho$ to the incompressibility condition (e.g. density, that stems from the continuity equation [2]), which are suitable to describe a large number of real cases with very small Reynolds number (Re<<1) where nonlinear aspects are absolutely not negligible but a strong compressibility constraint is not binding [Ref3] (a typical chaotic behaviour); so that we can put constant the dynamic viscosity and the local Cauchy problem [4] reduces to the weakly compressible limit of the Navier-Stokes initial-value problem,

$$\frac{\partial \vec{u}}{\partial t} = -\left[ \left( \rho^{-1}\nabla p + \nabla \cdot (\vec{u}\vec{u}) \right) + \left( \vec{u} + \rho^{-1}\nabla(\lambda + 2\mu) \right)\frac{d}{dt}\ln\varepsilon_\rho + \rho^{-1}\mu\nabla \wedge \left( \nabla \wedge \vec{u} \right) \right] + \vec{f}$$

$$\nabla \cdot \vec{u} = -\frac{d}{dt}\ln\varepsilon_\rho, \quad \varepsilon_\rho(x,y,z;t) \approx 0 \quad \forall (x,y,z;t) \in \Omega' \times [0,+\infty) \subset \mathbb{R}^4 \wedge \Omega' \supset \Omega \qquad [5]$$

$$\vec{u}(x,y,z;0) = \vec{h}(x,y,z) \quad \forall (x,y,z) \in \Omega$$

where the second relationship holds for time-dependent perturbations close to zero in the compressibility (density fluctuations) constraint. It is crucial to observe that we cannot completely avoid the nonlinearity only by considering incompressible fluids, such that

$$\frac{\partial u}{\partial x} + \frac{\partial v}{\partial y} + \frac{\partial w}{\partial z} = 0 \quad [6]$$

because the presence of a residual nonlinearity inside the tensorial term

$$\nabla \cdot (\vec{u}\vec{u})$$

so that almost always we have to solve a nonlinear differential problem in order to compute the right fluid behaviour.
Moreover, the last one can establish also a good reference model to analyze local turbulent frames in a fluid when we put our attention on slightly more large Reynolds number (Re≈1) like in several cases coming from fluidics of the mesoscales.

The fundamental features that point to the usefulness of the system [5] to reproduce quite well the dynamics in a wide range of fluid problems can be summarized by using the *Cauchy-Kowalewski Theorem*, which asserts essentially that '*looking at the initial-value problem

$$\frac{\partial u_i}{\partial t} = A_i\left(u_j, \frac{\partial u_j}{\partial x_k}, \frac{\partial^2 u_j}{\partial x_k \partial x_p}, \ldots\right) \quad [7]$$

$$u_i(\vec{x};0) = h_i(\vec{x}), \quad \vec{x} = (x_1, \ldots, x_r)$$

*if the differential operators $A_i$ are analytic, then there exists one and only one local solution of [7] for any analytic choice of the initial values h'* [Ref4].
This means that we need to focus our attention on all those solutions of [7] which are 'at least' strictly in respect of the analiticity of the initial values $h_i$, regardless to particular smoothness criteria in giving their definition, due that a wide class of fluid dynamics is well bounded under the only analytic constraint.

**The physical problem**

Certainly we know what problems arise when we move to finite-difference analysis of a certain PDE scheme. Speaking about the reason of the present work, due to the general complexity of the system [7], which in most cases, like in [5], can be viewed as a quasi-linear system, the question related to the definition of a *well-set initial-value problem in the sense of Hadamard* is not trivial [Ref5,Ref6] and it is usually hard to solve it analitically: in fact, keeping in care all the possible values of the 'input parameters' **($a_1, \ldots, a_n$)**, it means to look for a way to bound all these values by localizing a certain physically reasonable sub-set **Π** in which any possible choice of **($a_1, \ldots, a_n$)** leads to an acceptable behaviour in the local dynamics of the fluid, where 'input parameters' stems to identify all the variables from which the initial value functions **h** depend,

$$\frac{\partial u_i}{\partial t} = A_i\left(u_j, \frac{\partial u_j}{\partial x_k}, \frac{\partial^2 u_j}{\partial x_k \partial x_p}, \ldots\right) \quad [8]$$

$$u_i(\vec{x};0) = [h_i(a_1, \ldots, a_n)](\vec{x}), \quad \vec{x} = (x_1, \ldots, x_r)$$

Putting in light the real nature in the *modulation* of the general solution in [8] due to **($a_1, \ldots, a_n$)**, very often we encounter the need to separate the physical parameters from the unphysical ones (that are not always distinguishable and in most cases some of them are implicitly hidden in the related finite-difference model), in order to discriminate amongst them and understand 'when' we can retain the physical meaning of a certain dynamical behaviour in our problem.
In any case, the general solution in [8] will deliver something depending on the set **($a_1, \ldots, a_n$)** as it follows,

$$u_i(\vec{x};t) = [H_i(a_1, \ldots, a_n)](\vec{x};t) \quad [9]$$

and it is clear how the fundamental question about *well-posedness* of [8] extends to the very important class of the *steady-state problems*, where undoubtly it is reasonable to assert that solutions do not depend on time by looking at large time scales, that means

$$u_i(\vec{x};t) \underset{t\to\infty}{\to} \left[\tilde{H}_i(a_1,\ldots,a_n)\right](\vec{x}) \quad [10]$$

in which $u_i$ tends to some specific steady spatial distribution modulated implicitly by the input set **($a_1$,...,$a_n$)**.
Why are these steady solutions so important in hydrodynamics finite-difference analysis?
This special sub-domain of integrals of the system [8] seems to approach a wide class of physical solutions in which we are interested, because both the velocity and pressure fields in the steady dynamics are *observables* and the fluid scenario is quite often reproducible from the experimental point of view (which seems very hard, for instance, in turbulence at high Re). Neverthless, there are several well known steady *benchmark tests* [Ref7] which can be used to determine *a priori* the right steady behaviour for an evolution model like [4] and we can consider them very useful to modulate correctly the steady solution by setting duely the parameters **($a_1$,...,$a_n$)** inside, *which is just a way to define smoothness criteria for the initial values in the fluid domain of interest*.
In such a case, the evolution bounding is obviously garanteed and it reflects automatically on the *well-posedness* of the related Cauchy problem.
Let us observe that when we look for such a class **Ξ** of solutions, then the problem [8] reduces clearly to the following steady PDE,

$$A_i\left(u_j, \frac{\partial u_j}{\partial x_k}, \frac{\partial^2 u_j}{\partial x_k \partial x_p}, \ldots\right) = 0 \quad [11]$$

with general solution, as stated above, given by

$$\tilde{u}_i(\vec{x}) = \left[\tilde{H}_i(a_1,\ldots,a_n)\right](\vec{x}), \quad \forall (a_1,\ldots,a_n) \in \Pi$$

**Numerical solvers**

Discrete-time integration procedures work properly only with very small time stepping in order to reduce the numerical error that necessarily arises in every discretization scheme of a given order. Indipendently how much they are refined, being far from being analytic solvers, it is well known that every computational scheme is subject to fall in crash when a possible perturbation (numerical) introduces inside the evolution; it does not matter how small the time step we choose.

Perturbation propagates through the dynamics, growing sometime up to diverge in a few steps, hiding irremediably the physical behaviour. Moreover, several nonlinear integration schemes (deterministic or stochastic) return a non-dissipative dynamics which does not preserve conservation of the energy in the overall fluid domain (the following figure is drawn from [Ref8]),

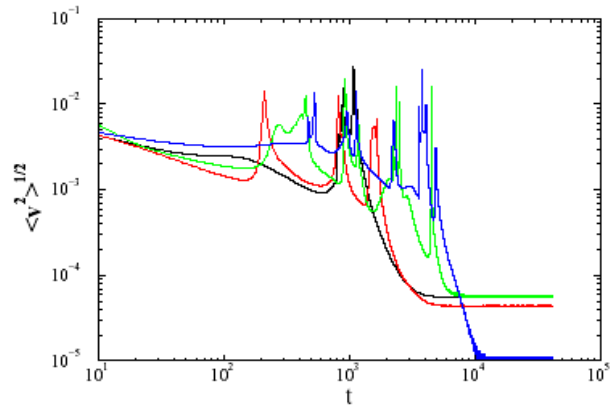

Figure 7. The time evolution of the root-mean-square velocity of four individual runs of the numerical simulations for different initial conditions. The applied stress value in all cases is $\sigma = 0.0075$. The curves are depicted in a double logarithmic scale to emphasize the intermittent bursts characteristic of the creep dislocation dynamics around the yield threshold.

(M. Carmen Miguel, José S. Andrade Jr., Stefano Zapperi, *Deblocking of interacting particle assemblies: from pinning to jamming,* Brazilian Journal of Physics, Vol. 33, No. 3, Sept., 2003)

In order to avoid such unwanted effects, it is undoubtly useful to put attention in reducing the related sources of numerical errors by a strong simplification of the finite-difference architecture and by increasing duely the order of the integrator; by the way, another source of error which is unphysical is given, as stated above, by ill-posedness aspects of the initial-value problem [8], so that, in order to recognize such an error and discriminate amongst all the possible sources of unphysical behaviours, we have necessarily imposing on the fluid velocity field appropriate initial conditions which make the related Cauchy problem well-posed without any doubt.
Surely we can modulate the initial values of the fluid velocity field by the ensemble of the simulation parameters **($a_1,...,a_n$)** which clearly hold for well-posedness in the underlying initial problem, but this is not possible *a priori* due the considerable wideness of the range in which they can variate; this means obviously that well-posedness is not a trivial question when we trust that our model reproduces exactly the dynamics of the physical phenomenon we want investigate and we have no way to decide what is better to do first, which is not banal.

**Bounding *well-posedness* – the method**

As stated above, the only thing we have to ensure is the strict analiticity in [8] of the initial values $h_i$ which are modulated by the parameters **(a₁,...,aₙ)**; it is possible to garantee this condition when there exists for sure some bounded solution of the given fluid model that, as showed before, must belong to the class of the steady solutions to be useful in determining the sub-set **Π** for which any possible solution of [8] is bounded and depends continuosly on the initial data [Ref5].

Therefore, in order to check for such a sub-set of parameters, we have to proceed by following three basic steps:

1 - check for a benchmark solution by looking at the sub-class of the well known steady behaviours of [5] ;

2 - reverse the model in time by using an appropriate change of variable **g(t)** such that *any possible initial-value problem [8] turns into a steady-state problem,*

$$\frac{\partial g}{\partial t}(t) < 0 \quad \forall t \in (-\infty, +\infty)$$

in which the previous condition on **g** provides that 'times flow continuosly moving with opposite verses';

3 - reverse the steady-state to a *new Cauchy problem* by using both the steady fluid velocity and pressure profile to garantee its *well-posedness* under the induced bounding of the time reversal solution.

Let us observe that the relationship between the time reversal solution and the time direct solution stems for any reasonable change of the time variable which must be necessarily a bijection (because the uniqueness of the direct evolution that follows from the Cauchy-Kowalewski theorem),

$$\frac{\partial \tilde{u}_i}{\partial t'}(\vec{x}; t') = \frac{\partial u_i}{\partial t}(\vec{x}; t) \frac{\partial g^{-1}}{\partial t'}(t') \quad [12]$$

It is clear that this procedural sequence states properly a method to define a new well-set initial-value problem (any possible steady state comes from some physically reasonable behaviour) which can permit us to determine the well-posedness constraints for **(a₁,...,aₙ)** in [8] by using the uniqueness (which is garanteed from the step '1') of the time reversal solution,

$$\frac{\partial \tilde{u}_i}{\partial t'} = A'_i\left(\tilde{u}_j, \frac{\partial \tilde{u}_j}{\partial x_k}, \frac{\partial^2 \tilde{u}_j}{\partial x_k \partial x_p}, \ldots\right) \quad [13]$$

$$\tilde{u}_i(\vec{x}; 0) = \left[\tilde{H}_i(a_1,\ldots,a_n)\right](\vec{x}), \quad \forall (a_1,\ldots,a_n) \in \Pi$$

where the new differential $A'_i$ operator is

$$A'_i = \frac{\partial g^{-1}}{\partial t'} A_i$$

Looking at the self-consistence of the present method, since our fundamental position inside step '2', which reads for

$$\partial_{t'} \tilde{u}_i \left(\vec{x}; +\infty\right) = 0 \quad [14]$$

the relationship [12] provide us a way to define criteria we need to impose well-posedness of the problem [8],

$$\partial_t u_i(\vec{x};t) \, \partial_{t'} g^{-1}(t') \underset{t' \to +\infty}{=} 0 \quad [15]$$

where the only constraint on $g^{-1}$ is given implictly inside the same step,

$$\frac{\partial g^{-1}}{\partial t'}(t') < 0 \quad \forall t' \in (-\infty, +\infty)$$

so that, exploring all the possible asyntotic choices for $\partial_{t'} g^{-1}$ in [15], the problem [14] can be stated *satisfied* only when any possible solution of the direct problem delivers

$$\partial_t u_i(\vec{x};t) \underset{t' \to +\infty}{=} o\left(\left[\partial_{t'} g^{-1}(t')\right]^{-1}\right) \quad [16]$$

In this view, we have to focus our attention on something that arises at this point as a natural consideration:

> *'the inverse problem [13] togheter with the steady constraint [14] state a boundary problem which is undoubtly well-posed'*

which means that a solution of the initial-value problem [13] is bounded only when we impose [14] in $\Pi$ in place of the direct problem [8],

$$\tilde{u}_i\left(\vec{x}; +\infty\right) = [h_i(a_1, \ldots, a_n)](\vec{x}) \quad \forall (a_1, \ldots, a_n) \in \Pi$$

**Decomposability benchmark**

In this section we want provide a clear example about a simple practical application of the method that we have treated above.
Hagen-Poiseuille flow (see following figure), which stems for a viscous fluid flow in a channel **Ω** made of two infinite parallel plates divided by a distance **h** (the amplitude of the channel), is a suitable benchmark fluid scenario whose steady velocity profile is well defined in the laminar regime (Re<<1),

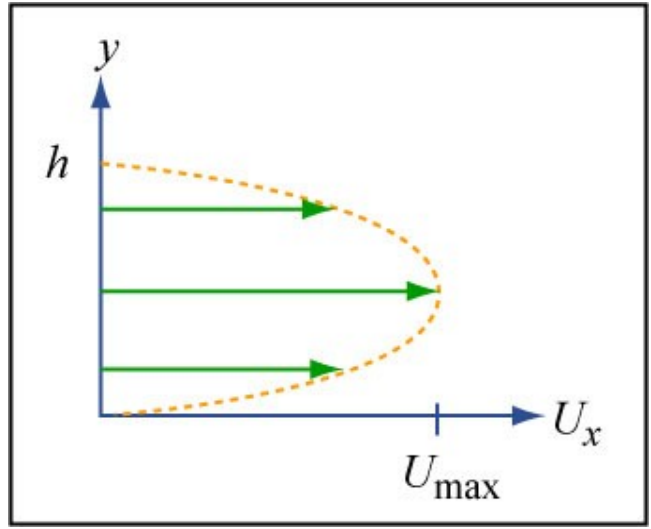

Looking at the incompressible limit

$$\nabla \cdot \vec{u} = 0$$

it is solution of [4] pertaining either the steady ensemble **Ξ** and the related symmetry conditions without external forces,

$$\nabla p + \mu \nabla \wedge \left( \nabla \wedge \vec{u} \right) = \vec{0}$$

in which the gradient of the dynamic pressure does not depend on time (we put constant the difference between outflow and inflow pressures which keeps the fluid moving); in such a case the general solution reads for

$$\vec{u}(\vec{x};+\infty) = u_x(\vec{x};+\infty)\vec{i} = \frac{1}{2\mu}\frac{\partial p}{\partial x}(y^2 - yh)\vec{i} \quad [17]$$

As stated in the previous section, the inverse problem is given by a suitable change of variable **g(t)** in the [8]-related direct problem togheter with the initial values coming by reversing [17] in time; this sets the following well-posed boundary problem,

$$\frac{\partial \tilde{u}_x}{\partial t'} t'^2 + \mu \frac{\partial^2 \tilde{u}_x}{\partial y^2} = \frac{\partial \tilde{p}}{\partial x} \qquad [18]$$

$$\tilde{u}_x(\vec{x};0) = \frac{1}{2\mu}\frac{\partial p}{\partial x}(y^2 - yh) \quad \forall \vec{x} \in \Omega$$

$$\partial_{t'} \tilde{u}_x(\vec{x};+\infty) = 0$$

where we have choosed the not analytic change

$$g(t) = \frac{1}{t} = t'$$

Due either the laminar regime and the simmetry conditions of the Hagen-Poiseuille flow, to solve the differential equation in [18] we recognize that the general direct solution is decomposable in such a manner,

$$u_x(\vec{x};t) = j(t)\, u_x(\vec{x};+\infty)$$

which stems for

$$\tilde{u}_x(\vec{x};t') = \tilde{j}(t')\, \tilde{u}_x(\vec{x};0)$$

in the inverse problem [18], so that it delivers the new 2nd-order linear problem

$$\tilde{u}_x(\vec{x};0)\frac{\partial \tilde{j}}{\partial t'} t'^2 + \mu \frac{\partial^2 \tilde{u}_x}{\partial y^2}(\vec{x};0)\, \tilde{j} = \frac{\partial \tilde{p}}{\partial x} \qquad [19]$$

whose steady solution is

$$\tilde{u}_x(\vec{x};+\infty) = \tilde{u}_x(\vec{x};0) + \left[\tilde{u}_x(\vec{x};\overline{t'}) - \tilde{u}_x(\vec{x};0)\right] \exp\left(\frac{-2\mu}{(y^2 - yh)\overline{t'}}\right) \qquad [20]$$

$$\forall \overline{t'} \in (0,+\infty)$$

It is now clear that to set well-posedness for the direct problem in order to avoid unphysical behaviours in the evolution dynamics, we must impose any possible choice of the initial data given inversely by [20] such that the velocity variation inside holds for

$$\tilde{u}_x\left(\vec{x};\overline{t'}\right) - \tilde{u}_x\left(\vec{x};0\right) \underset{\overline{t'} \to 0^+}{\in} O\left[\exp\left(\frac{2\mu}{(y^2 - yh)\overline{t'}}\right)\right]$$

as for a fluid which flows into the channel we have

$$\frac{2\mu}{y^2 - yh} < 0$$

**Conclusions**

Under the evidence about the practical need to impose some input parameters **(a_1,...,a_n)** in the initial-value problem for the weakly compressible Navier-Stokes equations [5] we want to simulate numerically, supposing the *existence* of a non empty sub-set **Ξ** of solutions for the related steady problem,

$$\nabla \cdot \vec{u}\vec{u} + \rho^{-1}\left[\nabla p + \mu \nabla \wedge \left(\nabla \wedge \vec{u}\right)\right] - \vec{f} = \vec{0}$$

we have briefly discussed on the opportunity to approach well-posedness, in the sense of Hadamard, by establishing a *new point of view* for the Cauchy problem that arises, based on the fundamental ascertainment that 'every solution of [5] which admits a steady behaviour must have necessarily a bounded evolution'; this states that it is possible in any case defining a well-posed *inverse problem* [13], by a suitable change of the time variable in the direct problem, which must evolve inversely to some steady state for an adequate choiche **Π** of the above parameters, imposing in such a way either the existence and bounding of the direct evolution which automatically reflects on the well-posedness of the direct problem.
Let us finally observe that the supposed steady problem [14] togheter with the problem [13] result in a well-set *boundary problem* for any choiche of **(a_1,...,a_n)** in the sub-set **Π**.

**Aknowledgements**

I want lovely thank my family in the person of my wife, Tiziana Longo, to have had a lot of patience in hearing all the questions about the present work and even to have discussed with me some fascinating aspects of the pure mathematical issues inside.

13**Reference works**

[Ref0] - Peter Constantin, Ciprian Foias, *Navier-Stokes equations*, Chicago lectures in Mathematics, University of Chicago Press, 1989

[Ref1] - Touvia Miloh, *Mathematical approaches in hydrodynamics*, SIAM, 1991

[Ref2] - Batchelor G.K., *An Introduction to Fluid Dynamics*, Cambridge University Press, 1967

[Ref3] - Hunana P., Zank G.P., Shaikh D., *Nearly incompressible fluids: hydrodynamics and large scale inhomogeneity*, Phys. Rev. E Stat. Nonlin. Soft Matter Phys., 74 (2 Pt 2): 026302, Aug, 2006

[Ref4] - Garrett Birkhoff, *Classification of partial differential equations*, J. Soc. Industr. Appl. Math., Vol. 2, No. 1, March, 1954

[Ref5] - V. Ya. Arsenin, *On ill-posed problems*, Russian Maths. Surveys, pp.93-107, 31:6, 1976

[Ref6] - Daniel D. Joseph, Jean Claude Saut, *Short-wave instabilities and ill-posed initial-value problems*, Theoret. Comput. Fluid Dynamics, Vol. 1:pp.191-227, 1990

[Ref7] - Drazin P.G., Reid W.H., *Hydrodynamic Stability, second edition*, Cambridge University Press, ISBN: 9780521525411, Sept. 20, 2004

[Ref8] – M. Carmen Miguel, José S. Andrade Jr., Stefano Zapperi, *Deblocking of interacting particle assemblies: from pinning to jamming,* Brazilian Journal of Physics, Vol. 33, No. 3, Sept., 2003

[Ref9] - Charles L. Fefferman, *Existence and smoothness of the Navier-Stokes equation*, in the Clay Mathematics Institute's list of prize problems---

Personal contacts:

M.Sc. Michele Romeo
Lehrstuhl für Aerodynamik
Technische Universität München
Boltzmannstr.15
85748 Garching
Germany

michele.romeo.mr@gmail.com

---